\RequirePackage{fix-cm}
\documentclass[twocolumn]{svjour3}          
\smartqed  
\usepackage{graphicx}
\usepackage{subfigure}
\usepackage{tabu}                      
\usepackage{booktabs}                  
\usepackage{hyperref}
\begin{document}

\title{Camera Travel for Immersive Colonography
}


\author{Soraia F. Paulo         \and
        Daniel Medeiros         \and
        Pedro B. Borges         \and
        Joaquim Jorge           \and
        Daniel Simões Lopes
}


\institute{Soraia F. Paulo, Pedro B. Borges, Joaquim Jorge, Daniel Simões Lopes \at
              INESC-ID Lisboa, Instituto Superior Técnico, ULisboa \\
              Av. Prof. Dr. Cavaco Silva, 2744-016, Porto Salvo, Portugal\\
              \email{daniel.lopes@inesc-id.pt}
           \and
           Daniel Medeiros \at
              CMIC, Victoria University of Wellington
}

\date{Received: date / Accepted: date}

\maketitle

\begin{abstract}
Immersive Colonography allows medical professionals to navigate inside the intricate tubular geometries of subject-specific 3D colon images using Virtual Reality displays. Typically, camera travel is performed via Fly-Through or Fly-Over techniques that enable semi-automatic traveling through a constrained, well-defined path at user controlled speeds. However, Fly-Through is known to limit the visibility of lesions located behind or inside haustral folds, while Fly-Over requires splitting the entire colon visualization into two specific halves. In this paper, we study the effect of immersive Fly-Through and Fly-Over techniques on lesion detection, and introduce a 
camera travel technique that maintains a fixed camera orientation throughout the entire medial axis path. While these techniques have been studied in non-VR desktop environments, their performance is yet not well understood in VR setups. We performed a comparative study to ascertain which camera travel technique is more appropriate for constrained path navigation in Immersive Colonography. To this end, we asked 18 participants to navigate inside a 3D colon to find specific marks. Our results suggest that the Fly-Over technique may lead to enhanced lesion detection at the cost of higher task completion times, while the Fly-Through method may offer a more balanced trade-off between both speed and effectiveness, whereas the fixed camera orientation technique provided seemingly inferior performance results. Our study further provides design guidelines and informs future work.

\keywords{Virtual Reality \and Colonography \and Navigation \and Medical Imagery \and Human-Centered Computing}

\end{abstract}

\section{Introduction}
\label{intro}

Colorectal cancer (CRC) is the second leading cause of cancer-related death in the western world with an estimated 1.4 million new cases every year worldwide, half of which end in death~\cite{Ferlay2015}. Computed Tomography Colonography (CTC) is an imaging technique that has been widely adopted for colonic examination for diagnostic purposes. Still, the colon is an organ with several inflections and numerous colonic haustral folds along its extension, which make navigation inside CTC 3D models a hard task~\cite{yao2010reversible}. 

While analyzing CTC content, radiologists work in the standard workstation, i.e., desktop, monitor, mouse and keyboard. However, using a 2D display to analyze 3D structures can lead to missing information~\cite{mirhosseini2014}. As conventional systems rely on 2D input devices and stationary flat displays, users often struggle to obtain the desired camera position and orientation, which requires several cumulative rotations, hence, making it hard to perceive the colon structure in 3D. 
To visualize such anatomically complex data, the immersion and freedom of movement afforded by VR systems bare the promise to assist clinicians to improve 3D reading, namely enabling more expedite diagnoses~\cite{Schuchardt07}. 

Travel is considered the most basic and important component of the VR experience, which is responsible for changing the user's viewpoint position and rotation in a given direction~\cite{bowman1997travel}.
Due to the complexity of large virtual environments, certain authors apply travel techniques that rely on path planning, i.e. path-based or path-constrained travel. In this family of techniques the user follows a previously defined path, where users can still control speed, viewpoint direction and local deviation, so that they can locally explore the virtual environment ~\cite{elmqvist2006navigation,elmqvist2007tour}. 
Path-based travel can also be done automatically, where both the path and the movement are predefined in a way that creates a smooth navigation experience. Such features are welcomed for virtual endoscopy applications~\cite{bartz2005virtual,he2001reliable,huang2005teniae}.

Considering the complexity of the colon's structure, travel follows a semi-automatic procedure which relies on centerline estimation to constrain the direction of movement, while users are given control over speed. The most conventional way of CTC travel consists of the Fly-Through technique ~\cite{hong19953d}, where camera orientation follows the centerline's direction. Nonetheless, the use of VR could enable more natural means of travel by decoupling camera orientation from direction of movement, in the sense that relative orientation can differ from the centerline's direction. That is the case of the Fly-Over technique, where relative orientation is perpendicular to the centerline's direction~\cite{hassouna2006flyover}. Although these techniques are commonly used in conventional setups, they have yet to be fully investigated in VR settings. Our work focuses on camera travel as a key component on surveying and identifying pathological features in CTC datasets. 
The semi-automatic nature of the process combined with the abrupt movements caused by the complexity of the colon's structure, may cause unwanted side-effects due to the difference between camera orientation in the virtual world and the user's real orientation~\cite{Robinett1992}.
To overcome this issue, we propose a technique called Elevator, where camera orientation is changed in order to match the user's real orientation. Using an immersive colonoscopy prototype  ~\cite{lopes2018interaction} we studied camera control techniques and their effectiveness on comprehensive landmark identification in order to address the following question:
\emph{Which of the considered visualization techniques is the most suitable to navigate inside the 3D reconstructed model of the colon?}

\section{Related Work}
\label{relatedWork}

Navigation inside colon structures is a difficult and non-trivial task to perform. The Fly-Through technique has been widely adopted since it was first proposed by Lichan Hong~\cite{hong19953d}. Radiologists tend to prefer this type of visualization due to its similarities with conventional colonos-copy, which include dealing with the same type of limitations. While moving in a given direction, lesion visibility is limited to the
colorectal tissue that is exposed to the normal of the viewing camera, which may lead to missing significant lesions. In order to address this and reduce redundancy, colorectal flattening proposed mapping the colon's cylindrical surface to a rectangular surface to create a full virtual view of the colon~\cite{Haker2000NondistortingFM}. Nonetheless, flattening algorithms are prone to error and require additional training to understand such 2D representation of the colon~\cite{wang2015novel}. The unfolded cube projection proposed projecting all views of the colon on the inside of a cube using six camera normals that move together along the centerline~\cite{vos2003three}. Even though unfolding the cube enhances lesion visibility, scanning all sides of the cube is a time-consuming task. 
Fly-Over is another visualization technique that tries to solve Fly-Through's limitations. In this case, the colon is divided in two unique halves by the centerline, each with a virtual camera~\cite{hassouna2006flyover}. This enables perpendicular perspective, producing increased surface coverage (99\% of surface visibility in one direction vs 93\% in Fly-Through in two navigation directions) and equally good sensitivity. In spite of the Fly-Over's positive impact, current CTC softwares, such as the V3D-Colon\footnote{\textit{http://www.viatronix.com/ct-colonography.asp}}, syngo.CT Colonography\footnote{Siemens Healthineers, 2017.~\textit{syngo.CT Colonography}} only include Fly-Through, flattening and the unfolded cube visualization techniques, since the use of the Fly-Over is restricted for patenting reasons. Still, all of these techniques continue to suffer from the drawback of using a 2D interface to interact with a 3D model. 

Since its inception, Virtual Reality (VR) has found applications across the medical domain, namely in medical education~\cite{codd2011virtual}, surgical planning and training tasks \cite{de2011progress,vosburgh2013surgery,shanmugan2014virtual}. More recently, VR has also been applied to diagnosis~\cite{king2016immersive,sousa2017vrrrroom,wirth2018evaluation}, where being able to carefully examine large and complex image datasets is crucial to producing insightful and complete results. Thus, controlling viewing position and orientation in expedite yet precise manners could potentially affect significant medical decisions.

Considering the advantages of VR under the scope of diagnostic imaging, especially improved camera control, freedom of movement, 3D perception and enhanced scale, two groups started to explore the immersive CTC experience. Mirhosseini et al. investigated a CAVE (Cave Automatic Virtual Environment) in which the gastrointestinal walls were projected onto the room walls~\cite{mirhosseini2014}. In spite of suggesting potential improvement in terms of reducing examination time and enhancing accuracy, this type of setup would be unrealistic in a real clinical setting. More recently, Mirhosseini et al. proposed an immersive CTC system which leverages the advantages afforded by VR to improve lesion detection~\cite{mirhosseini2019immersive}, while still relying on 2D interaction techniques. Similarly, Randall et al. explored an immersive VR prototype, obtaining promising feedback regarding overall faster diagnosis~\cite{randall2015oculus}. However, none of these works focus on camera travel, nor explore a camera travel technique other than the Fly-Through.
%

\section{Immersive Navigation of a 3D Virtual Colon}
\label{approach}

An interactive VR system 
was developed to assist 3D immersive navigation of subject-specific colon models enabling users to travel via Fly-Through, Fly-Over or Elevator camera modes~\cite{lopes2018interaction}.

\subsection{3D Data}

\begin{figure}[b]
    \centering
    \includegraphics[width=\columnwidth]{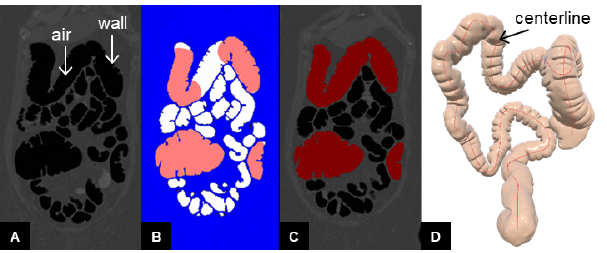}
    \caption{3D colon reconstruction: (A) original CTC image; (B) global threshold image with two active contours (red); (C) segmented colon overlapped with CTC image; (D) reconstructed 3D model with centerline}
    \label{fig:colonSegmentation}
\end{figure}

A single CTC dataset from \textit{The Cancer Imaging Archive} \cite{tciaCTC} was considered (subject ID \textit{CTC-3105759107}). 
The 3D model was reconstructed using the image-based geometric modelling pipeline that is composed by freeware tools~\cite{ribeiro20093} (\autoref{fig:colonSegmentation}).

The high contrast between luminal space (air: black) and colon luminal surface (colon wall: light grey) facilitates 3D reconstruction (\autoref{fig:colonSegmentation}(A)). Firstly, the 3D colon structure is segmented using the active contours method based on region competition (\autoref{fig:colonSegmentation}(B-C)), which depends on the intensity values estimated via a global threshold filter (ITK-SNAP 3.6). Secondly, a 3-D surface mesh of the segmented data is generated using marching cubes. Thirdly, undesired mesh artifacts were attenuated through a cycle of smoothing and decimating operations (ParaView 5.3.0) and exported into a *.ply (ASCII) file. Finally, the mesh file was converted to *.obj (Blender 2.78) and imported into Unity. To compute the 3D centerline of the colon mesh, we used the algorithm proposed by Tagliasacchi et al.~\cite{tagliasacchi2012mean} which solves the 3D mesh skeletonization problem by resorting on mean curvature flow (\autoref{fig:colonSegmentation}(D)). 

\subsection{Immersive Camera Travel Techniques}

\begin{figure}[t]
\centering
\includegraphics[width=.65\columnwidth]{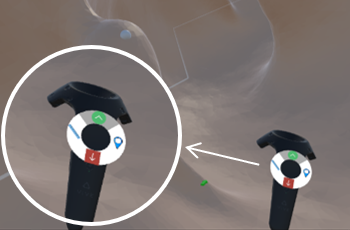}
\caption{Touchpad controller indicating the direction of movement: pressing the green arrow button the user moves forward, while pressing the red arrow the user moves backwards.}
\label{fig:interaction}
\end{figure}

We considered three camera travel techniques which allowed users to inspect the colon model and navigate inside the luminal space: Fly-Through, Fly-Over and Elevator. All techniques follow a predefined centerline, i.e., follow the same path and use the same input (touchpad) to indicate the direction of movement (forward or backwards) at a constant speed (\autoref{fig:interaction}). 

\begin{figure}[h!]
\centering
\subfigure[Fly-Through]{
\includegraphics[width=\columnwidth]{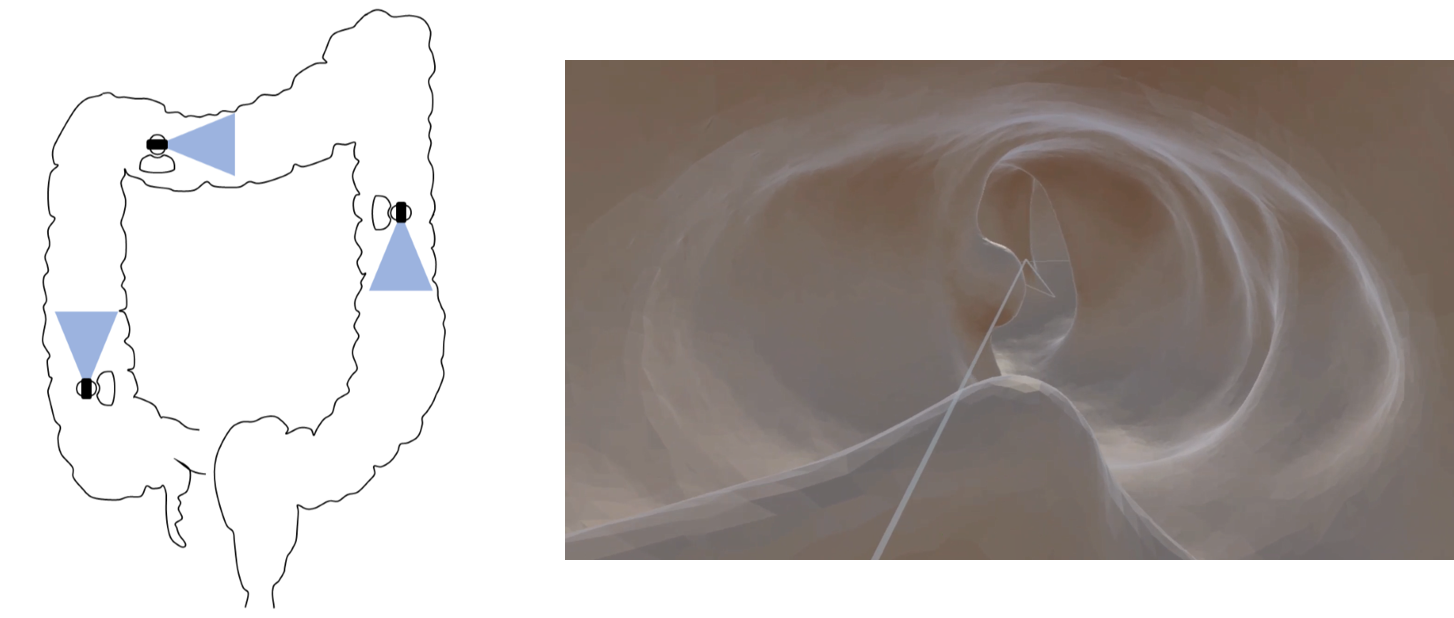}%
\label{fig:flyThrough}
}
\subfigure[Fly-Over]{
\includegraphics[width=\columnwidth]{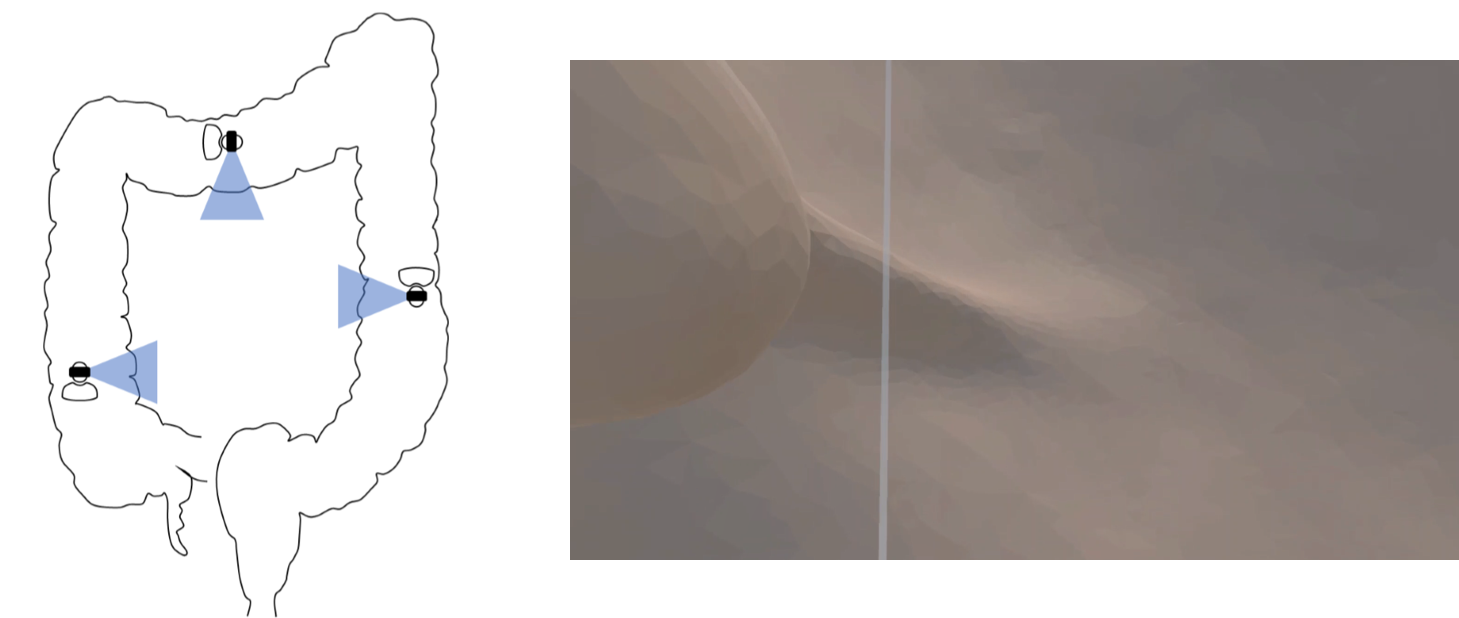}%
\label{fig:flyOver}
}
\subfigure[Elevator]{
\includegraphics[width=\columnwidth]{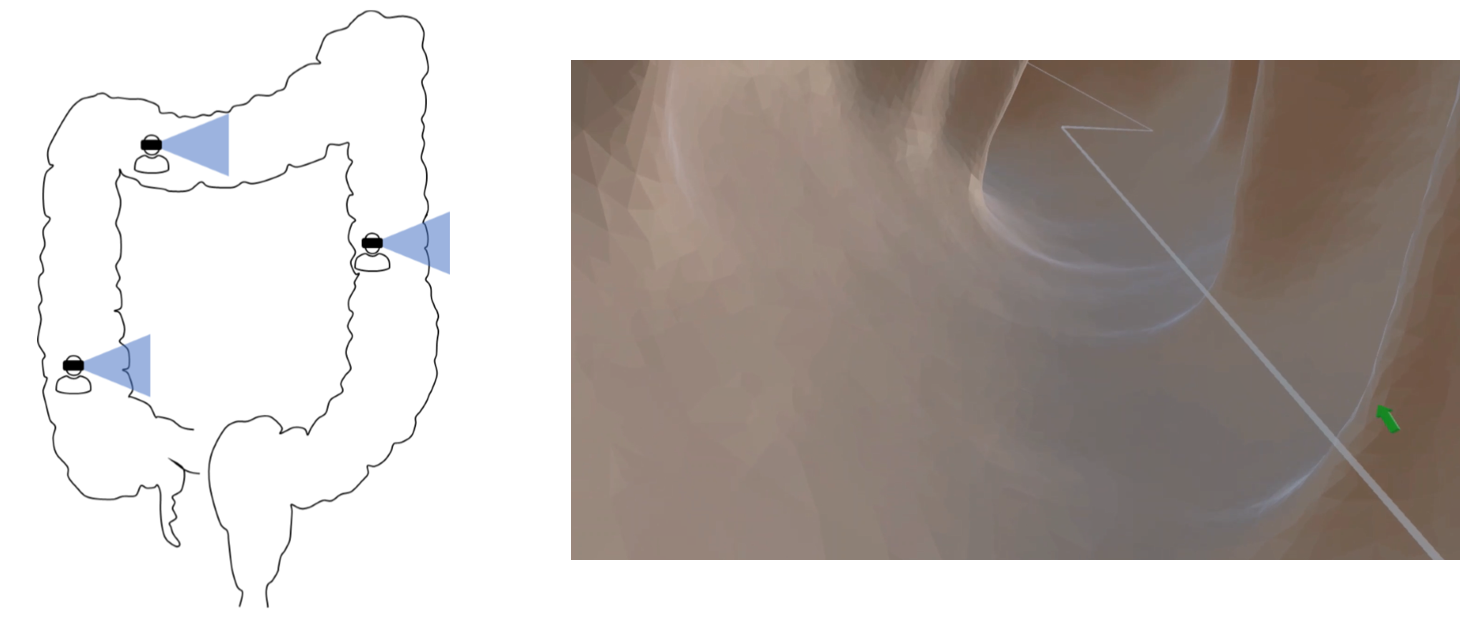}%
\label{fig:elevator}
}
\caption{Camera orientation schematics (left) and viewpoints (right) during (a) Fly-Through, (b) Fly-Over and (c) Elevator techniques. }
\label{fig:camOrientation}
\end{figure}

Each technique differs on how the user’s orientation is represented within the virtual environment. 
Identical to conventional CTC, the Fly-Through will make the user feel inside a cave. In this technique, the virtual camera follows the path without the need for users to move their head. They can, however, move their heads to see what is behind, below or above them. User orientation follows the centerline's direction (\autoref{fig:camOrientation}(a)).
Differently from the traditional Fly-Over technique found in the literature~\cite{hassouna2006flyover} there is no need to split the colon in two halves and assign a virtual camera to each part. In this case, the inspection of the colon’s walls is done by users' head movement. The camera will automatically keep the perpendicular perspective in the eyes of the users, while they can move their heads to analyze their surroundings (\autoref{fig:camOrientation}(b)).
Finally, the Elevator technique does not change camera orientation, in order to match the user's real orientation (\autoref{fig:camOrientation}(c)). Ultimately, this could reduce cybersickness during the VR trip, at the cost of increasing users' chances of losing the sense of direction.

\section{Evaluation}
\label{evaluation}

We compared three different camera travel techniques in order to investigate their potential effects on efficiency and diagnosis accuracy during CTC navigation: Fly-Over, Fly-Through and Elevator.
We used both quantitative and qualitative metrics to assess the ease of use, usefulness, efficiency and efficacy of each technique. 
Efficiency was measured based on task completion time, as efficacy corresponded to the success rate, i.e. the percentage of specific marks that were correctly identified. Questionnaires were used to assess the subjective feeling of usefulness, ease of use and disorientation of all three techniques, as well as cybersickness~\cite{laviola2000discussion}. 

\subsection{Apparatus}

\begin{figure}[t]
\centering
\includegraphics[width=.9\columnwidth]{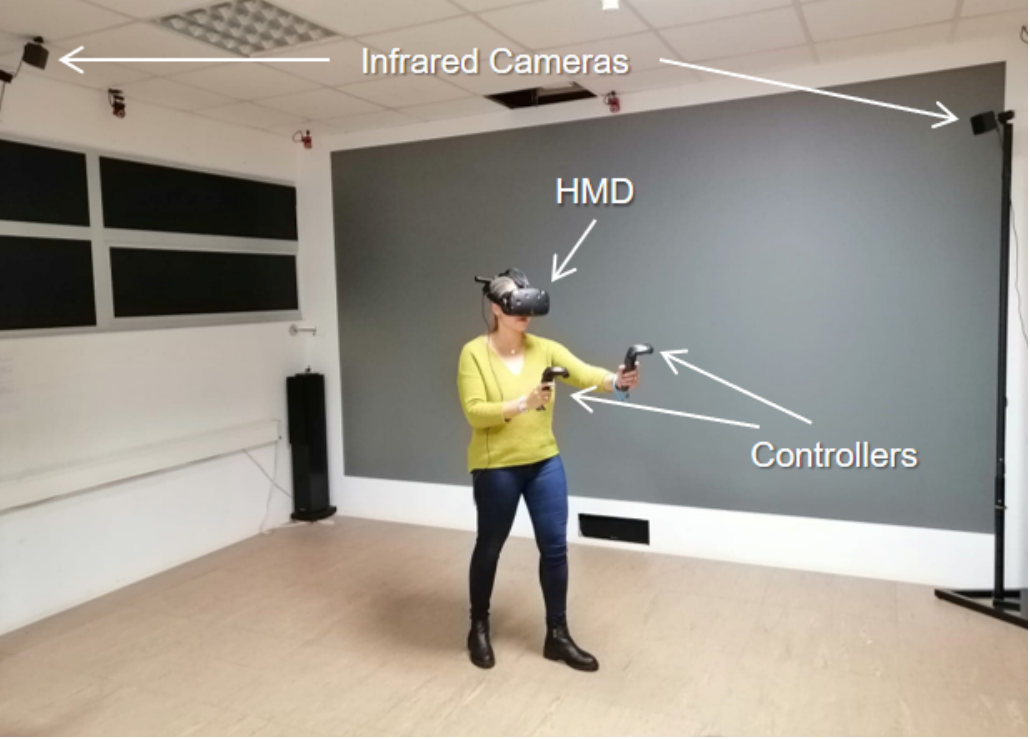}
\caption{VR setup of the immersive CTC interactive system.}
\label{fig:setup}
\end{figure}

Our setup relies on the off-the-shelf HTC Vive device. It consists of a binocular Head-Mounted Display, two game controllers and a Lighthouse Tracking System composed by two cameras with emitting pulsed infrared lasers that track all 6 degrees-of-freedom of head and handheld gear (\autoref{fig:setup}). 
The tracking system generates an acquisition volume that enables users to move freely within a 4.5x4.5x2.5 m\textsuperscript{3} space. 
User tests were performed on an Asus ROG G752VS Laptop with an Intel\textsuperscript\textregistered Core\texttrademark ~i7-6820HK Processor, 64GB RAM and NVIDIA GeForce GTX1070. The VR prototype runs at 60 frames per second.
All the code was developed in C\# using the SteamVR Plugin and Unity game engine (version 5.5.1f1).

\subsection{Participants}

Eighteen participants (13 male, 5 female) took part in our study, with ages between 18 and 25 years old (Mean = 21.94; Standard Deviation = 1.98). Most participants had an engineering background, namely Computer Science (38.89\%) and Biomedical Engineering (27.78\%). Most reported no previous experience in VR (66.6\%) or to use such systems less than once a month (27.78\%). One user reported to have claustrophobia.

\subsection{Methodology}

\begin{figure}[t]
\centering
\includegraphics[width=.6\columnwidth]{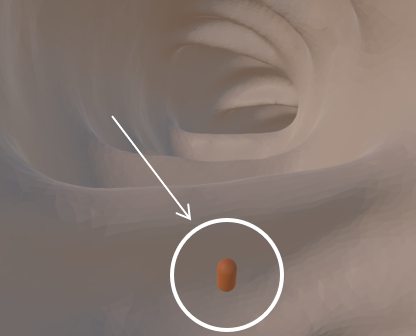}
\caption{Task performance: users were asked to find as many marks (orange) as they could while navigating inside the colon.}
\label{fig:prop}
\end{figure}

Participants were asked to complete a demographic questionnaire to survey their personal profile and previous experience regarding VR and medical tools.
This was followed by performing a training task with the technique they were assigned, to familiarize themselves both with the technique and the virtual environment.
After that, they performed the test task followed by a post-test questionnaire to assess qualitative metrics. The task consisted in finding specific marks, in the form of orange capsules (\autoref{fig:prop}), which were placed in both easy and hard to find locations in the colon, to simulate the visibility of real lesions. Users were oblivious to the total amount of marks (20 marks per technique) spread throughout the colon. Instead, they were asked to find as many as they could, until they felt they had found them all.
This procedure was repeated for all three techniques, which were assigned according to a balanced latin-squares arrangement to avoid learning effects.

\section{Results}

In this section we present the results from our statistical analysis to evaluate quantitative and qualitative measures regarding the three techniques tested.
For task completion time and success rate, a Shapiro-Wilk test revealed the data was not normally distributed. We thus applied a Friedman non-parametric test for multiple comparisons and Wilcoxon Signed-Ranks post-hoc tests with a Bonferroni correction, setting a significance level at $p<=0.017$. Such tests were also applied to Likert-scale data collected via questionnaires and cybersickness scores.

\begin{figure}[b]
    \centering
    \includegraphics[width=\columnwidth, trim=15 30 0 50, clip]{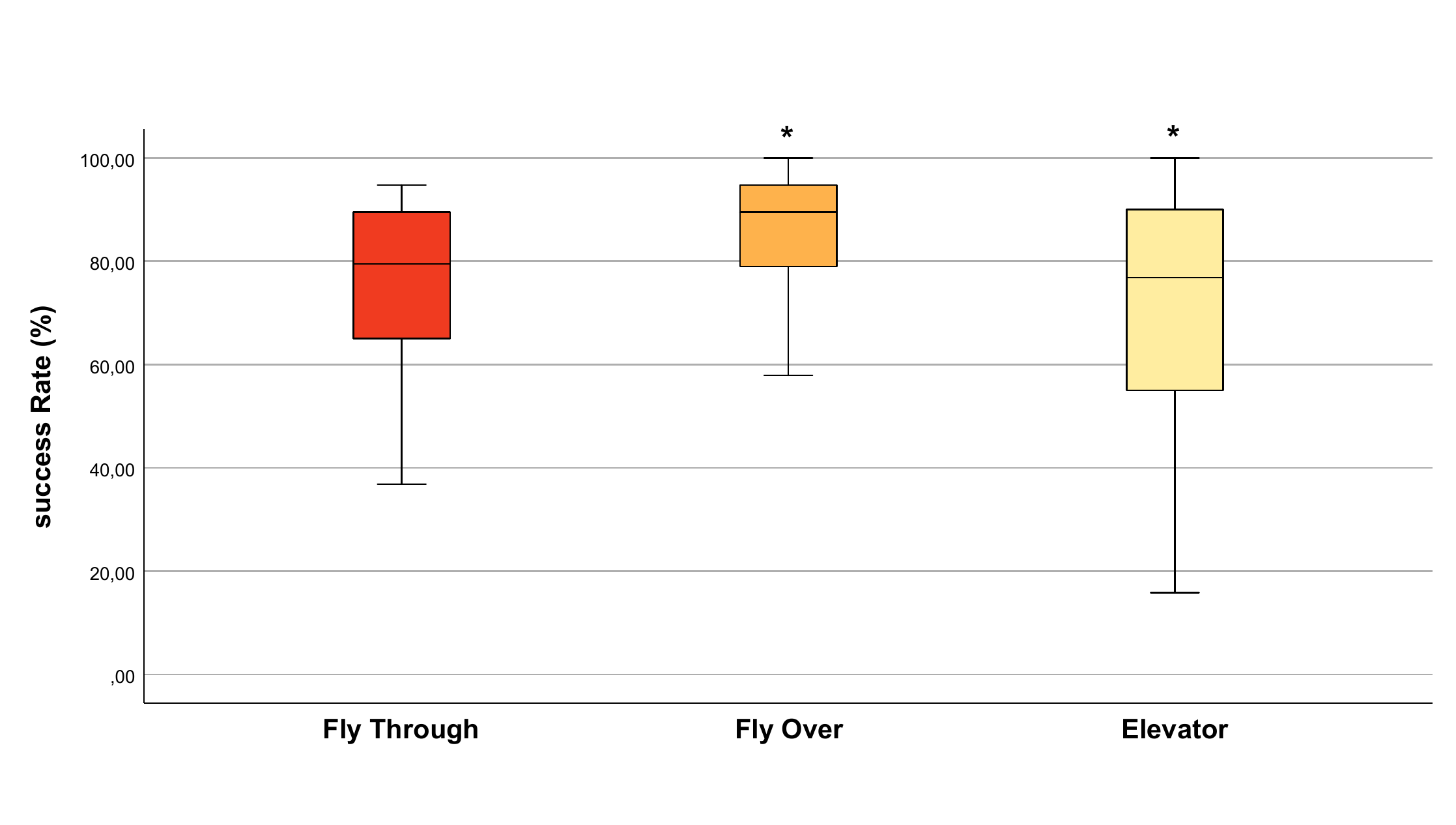}
    \caption{Success Rate for each condition: Fly-Through, Fly-Over and Elevator.* indicates statistical significance.}
    \label{fig:successRateBoxplot}
\end{figure}

There were 
significant differences in the success rate values depending on 
the camera travel technique used, $\chi^2(2) = 7.600$, $p = 0.022$. 
Median (Interquartile Range) values for success rate using the Fly-Through, Fly-Over and Elevator techniques were 79.47 (26.25), 89.47 (18.42) and 76.84 (36.77), respectively (\autoref{fig:successRateBoxplot}). Post-hoc analysis 
showed a statistically significant increase of the success rate between the Elevator and the Fly-Over technique ($Z=-2.386, p=0.017$). However, there were no statistically significant differences between the Fly-Through and Fly-Over techniques ($Z = -2.345, p = 0.019$), nor between the Fly-Through and the Elevator ($Z = -0.734, p =0.463$).   


\begin{figure}[t]
    \centering
    \includegraphics[width=\columnwidth, trim=15 30 0 0, clip]{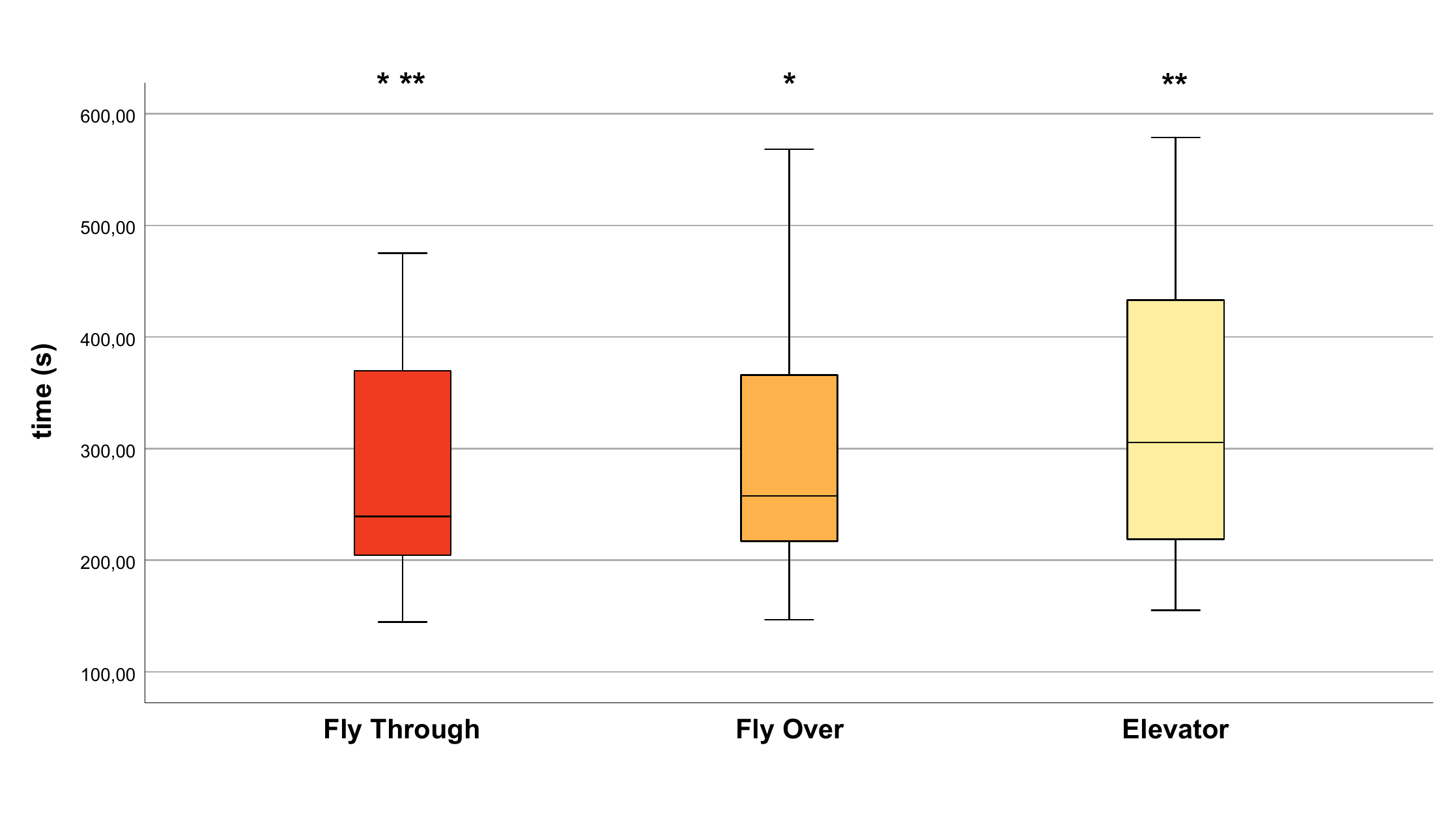}
    \caption{Task Completion Time for each condition: Fly-Through, Fly-Over and Elevator.* and ** indicate statistical significance.}
    \label{fig:timeBoxplot}
\end{figure}

Regarding task completion time, there were statistically significant differences depending on the camera travel technique 
used, $\chi^2(2) = 10.333$, $p = 0.006$. Mean (standard deviation) task completion time values for the Fly-Through, Fly-Over and Elevator techniques were 273.91 (100.05), 305.53 (124.38) and 322.13 (135.49), respectively (\autoref{fig:timeBoxplot}). Post-hoc analysis showed statistically significant decreases between Fly-Through and Fly-Over ($Z=-2.548, p=0.011$), as well as between the Fly-Through and the Elevator techniques ($Z = -2.548, p =0.011$). Still, there was no significant difference between the Fly-Over and the Elevator ($Z = -1.328, p = 0.184$). We also did not find significant differences between techniques regarding cybersickness scores. 

As 
for qualitative metrics (\autoref{tab:questionnaires}), we found statistical significance in the perceived usefulness of the navigation technique (Q1) ($\chi^2(2)=7.35$ $p=0.025$).
Notably, users found Fly-Through more useful than the Fly-Over technique ($Z=-2.588$ $p=0.01$).
We also found statistical significance regarding the ease of understanding the direction of movement (Q2) ($\chi^2(2)=9.529$ $p=0.009$), but with no significance between pairs 
after performing 
a Bonferroni correction.
Finally, results indicate statistically significant differences in 
perceived disorientation ($\chi^2(2)=11.111$ $p=0.004$).
In 
effect, users felt less disoriented by the Fly-Through technique when compared either to the Elevator ($Z=-2.541$ $p=0.011$) or the Fly-Over ($Z=-2.634$ $p=0.008$) methods. 
\begin{table}[t]
\caption{Summary of the questionnaires split by question and technique (Fly-Through (FO), Fly-Over (FO) and Elevator (EL)). Results are shown as Median (Interquartile Range).}
\resizebox{\columnwidth}{!}{
\begin{tabular}{@{}llll@{}}
\toprule
                                                                                                           & FT & FO & EL  \\ \midrule
Q1: Navigation was useful*                                                                                  & 6(1)       & 5(2)      & 6(2)      \\ \\
\begin{tabular}[c]{@{}l@{}}Q2: Direction of movement was easy to understand\end{tabular} & 6(0.25)    & 5(2)      & 5(2.25)   \\ \\
Q3: I was disoriented*                                                                                          & 1(1.5)     & 3.5(3.25) & 3(3.25)   \\ \\
Q4: It was easy to find the marks                                                                              & 5(2)       & 4.5(1.5)  & 5(2)      \\ \\
Q5: I felt that I found the same mark twice                                                                    & 2(2.25)    & 2.5(2.25) & 1.5(2.25) \\ \bottomrule
\end{tabular}
}
\label{tab:questionnaires}
\end{table}%

\section{Discussion}

Overall, Fly-Through has proven to be the best technique for immersive colonoscopy navigation in the user tests we conducted. Indeed, we found it to be the most efficient option, according to task completion times, besides being considered the most useful (Q1), easy to use (Q2) and less disorienting (Q3) by the subjects. Even though Fly-Over seemingly produced 
higher success rates, there were no statistical differences 
that could support its use over the Fly-Through, since 
the significant increase in task completion times would likely offset those gains. 
%
Additionally, users 
reported higher disorientation while using the Fly-Over technique. Such results may be attributed to the fact that 
Fly-Over had subjects 
facing the colon's walls most of the time. 
Orienting the camera at a direction perpendicular to displacement 
severely hampered their general perception of the tubular structure of the colon and the path they were following. This, combined with marks
located behind their 
backs, which forced subjects to inspect the structure in several directions to try and find them, ultimately caused their disorientation. Such results suggest that the Fly-Over technique may be improved by 
devising new means and interaction techniques for clinicians to visualize structures on their back without the need to physically turn.
By doing this they could combine both the observed effectiveness of the Fly-Over technique with more efficient means to 
support camera travel in immersive CTC navigation.
Finally, the Elevator technique was the least favored option 
for navigating the virtual environment. 
This may be due to the search strategy adopted by most, which had to change after each abrupt movement caused by the natural inflexions of the colon structure. That may also explain why this technique turned out to be the least efficient when compared to the other two approaches, as users required more time to adapt and adjust their orientation whenever the camera direction changed.

\section{Conclusions}

In this paper, we 
study Fly-Through and Fly-Over techniques in immersive VR CTC, 
in terms of efficiency, ease of use, usefulness and effectiveness. We also compared these to the Elevator, a novel technique in this domain that combines both camera techniques to make virtual orientation match the user's direction of movement throughout navigation. Our results show that Fly-Through is still the most efficient and easy to use technique for immersive VR CTC. The Elevator technique was found to be less effective and efficient than both Fly-Through and Fly-Over methods, but less disorienting than the Fly-Over approach.
This can be explained by the need to physically turn one's body to 
effectively scan the colon structure in all directions. Still, this limitation did not affect task effectiveness, as in the Fly-Over technique users were able to 
achieve higher success rates when finding specific marks along the colonic structure.
Thus, the Fly-Over would be the technique of choice in order to provide a more accurate analysis, and produce enhanced readings, as it helps people to identify lesions even in difficult-to-scan locations 
despite being a more time-consuming procedure. Indeed our experience suggests that each interaction technique could be useful in its own right, Fly-Through being most adequate to scan the colon in a quick preview, while Fly-Over would likely enable more reliable and comprehensive readings by clinicians. 

Still, our study had two main limitations. First, our experimental task only aims at reflecting the real clinical task to a certain extent, i.e. limited lesion visibility caused by the anatomical properties of the colon, while orange capsules may significantly differ from lesions such as polyps. Second, our participants had no clinical background, which may impact the selection of the ideal navigation technique to perform immersive VR CTC analysis. Future work will include validating such conclusions next to a group of medical professionals and the use of more generic flying to be able to generalize our results to a more broad area of cave-like structures.


\begin{acknowledgements}
This work was supported by Fundação para a Ciência e a Tecnologia through grants UIDB/50021/2020 and SFRH/BD/136212/2018, and by the Entrepreneurial University Program funded by TEC, New Zealand.
\end{acknowledgements}

%
\section*{Conflict of interest}

The authors declare that they have no conflict of interest.

\bibliographystyle{spmpsci}      
\bibliography{main}   

\begin{thebibliography}{10}
\providecommand{\url}[1]{{#1}}
\providecommand{\urlprefix}{URL }
\expandafter\ifx\csname urlstyle\endcsname\relax
  \providecommand{\doi}[1]{DOI~\discretionary{}{}{}#1}\else
  \providecommand{\doi}{DOI~\discretionary{}{}{}\begingroup
  \urlstyle{rm}\Url}\fi

\bibitem{Ferlay2015}
Ferlay, J., Soerjomataram, I., Dikshit, R., Eser, S., Mathers, C., Rebelo, M.,
  Parkin, D.M., Forman, D., Bray, F.: {Cancer incidence and mortality
  worldwide: Sources, methods and major patterns in GLOBOCAN 2012}.
\newblock International Journal of Cancer \textbf{136}(5), E359--E386 (2015).
\newblock \doi{10.1002/ijc.29210}

\bibitem{yao2010reversible}
Yao, J., Chowdhury, A.S., Aman, J., Summers, R.M.: Reversible projection
  technique for colon unfolding.
\newblock IEEE Transactions on Biomedical engineering \textbf{57}(12),
  2861--2869 (2010)

\bibitem{mirhosseini2014}
Mirhosseini, K., Sun, Q., Gurijala, K.C., Laha, B., Kaufman, A.E.: {Benefits of
  3D immersion for virtual colonoscopy}.
\newblock In: 2014 IEEE VIS International Workshop on 3DVis (3DVis), pp. 75--79
  (2014).
\newblock \doi{10.1109/3DVis.2014.7160105}

\bibitem{Schuchardt07}
Schuchardt, P., Bowman, D.A.: The benefits of immersion for spatial
  understanding of complex underground cave systems.
\newblock In: Proceedings of the 2007 ACM Symposium on Virtual Reality Software
  and Technology, VRST ’07, p. 121–124. Association for Computing
  Machinery, New York, NY, USA (2007).
\newblock \doi{10.1145/1315184.1315205}.
\newblock
  \urlprefix\url{https://doi-org.librarylink.uncc.edu/10.1145/1315184.1315205}

\bibitem{bowman1997travel}
Bowman, D.A., Koller, D., Hodges, L.F.: Travel in immersive virtual
  environments: An evaluation of viewpoint motion control techniques.
\newblock In: Proceedings of IEEE 1997 Annual International Symposium on
  Virtual Reality, pp. 45--52. IEEE (1997)

\bibitem{elmqvist2006navigation}
Elmqvist, N., Tsigas, P.: On navigation guidance for exploration of 3d
  environments.
\newblock G{\"o}teborg, Sweden  (2006)

\bibitem{elmqvist2007tour}
Elmqvist, N., Tudoreanu, M.E., Tsigas, P.: Tour generation for exploration of
  3d virtual environments.
\newblock In: Proceedings of the 2007 ACM Symposium on Virtual Reality Software
  and Technology, VRST ’07, p. 207–210. Association for Computing
  Machinery, New York, NY, USA (2007).
\newblock \doi{10.1145/1315184.1315224}

\bibitem{bartz2005virtual}
Bartz, D.: Virtual endoscopy in research and clinical practice.
\newblock Computer Graphics Forum \textbf{24}(1), 111--126 (2005).
\newblock \doi{10.1111/j.1467-8659.2005.00831.x}

\bibitem{he2001reliable}
He, T., Hong, L., Chen, D., Liang, Z.: Reliable path for virtual endoscopy:
  ensuring complete examination of human organs.
\newblock IEEE Transactions on Visualization and Computer Graphics
  \textbf{7}(4), 333--342 (2001).
\newblock \doi{10.1109/2945.965347}

\bibitem{huang2005teniae}
Huang, A., Roy, D., Franaszek, M., Summers, R.M.: Teniae coli guided navigation
  and registration for virtual colonoscopy.
\newblock In: IEEE Conference on Visualization, VIS 05, pp. 279--285. IEEE
  (2005).
\newblock \doi{10.1109/VISUAL.2005.1532806}

\bibitem{hong19953d}
Hong, L., Kaufman, A., Wei, Y.C., Viswambharan, A., Wax, M., Liang, Z.: 3d
  virtual colonoscopy.
\newblock In: Biomedical Visualization, 1995. Proceedings., pp. 26--32. IEEE
  (1995)

\bibitem{hassouna2006flyover}
Hassouna, M.S., Farag, A.A., Falk, R.: Virtual fly-over: A new visualization
  technique for virtual colonoscopy.
\newblock In: MICCAI 2006, pp. 381--388. Springer Berlin Heidelberg, Berlin,
  Heidelberg (2006)

\bibitem{Robinett1992}
Robinett, W., Holloway, R.: Implementation of flying, scaling and grabbing in
  virtual worlds.
\newblock In: Symposium On Interactive 3d Graphics, pp. 189--192. ACM Press
  (1992).
\newblock \doi{http://doi.acm.org/10.1145/147156.147201}

\bibitem{lopes2018interaction}
Lopes, D.S., Medeiros, D., Paulo, S.F., Borges, P.B., Nunes, V., Mascarenhas,
  V., Veiga, M., Jorge, J.A.: Interaction techniques for immersive ct
  colonography: A professional assessment.
\newblock In: MICCAI 2018, pp. 629--637. Springer (2018).
\newblock \doi{10.1007/978-3-030-00934-2_70}

\bibitem{Haker2000NondistortingFM}
Haker, S., Angenent, S.B., Tannenbaum, A.R., Kikinis, R.: Nondistorting
  flattening maps and the 3-d visualization of colon ct images.
\newblock IEEE Transactions on Medical Imaging \textbf{19}, 665--670 (2000)

\bibitem{wang2015novel}
Wang, H., Chen, Y., Li, L., Pan, H., Gu, X., Liang, Z.: A novel colon wall
  flattening model for computed tomographic colonography: method and
  validation.
\newblock Computer Methods in Biomechanics and Biomedical Engineering: Imaging
  \& Visualization \textbf{3}(4), 213--221 (2015)

\bibitem{vos2003three}
Vos, F.M., van Gelder, R.E., Serlie, I.W., Florie, J., Nio, C.Y., Glas, A.S.,
  Post, F.H., Truyen, R., Gerritsen, F.A., Stoker, J.: Three-dimensional
  display modes for ct colonography: conventional 3d virtual colonoscopy versus
  unfolded cube projection.
\newblock Radiology \textbf{228}(3), 878--885 (2003)

\bibitem{codd2011virtual}
Codd, A.M., Choudhury, B.: Virtual reality anatomy: Is it comparable with
  traditional methods in the teaching of human forearm musculoskeletal anatomy?
\newblock Anatomical sciences education \textbf{4}(3), 119--125 (2011)

\bibitem{de2011progress}
De~Visser, H., Watson, M.O., Salvado, O., Passenger, J.D.: Progress in virtual
  reality simulators for surgical training and certification.
\newblock Medical Journal of Australia \textbf{194}, S38--S40 (2011)

\bibitem{vosburgh2013surgery}
Vosburgh, K.G., Golby, A., Pieper, S.D.: Surgery, virtual reality, and the
  future.
\newblock Studies in health technology and informatics \textbf{184}, vii (2013)

\bibitem{shanmugan2014virtual}
Shanmugan, S., Leblanc, F., Senagore, A.J., Ellis, C.N., Stein, S.L., Khan, S.,
  Delaney, C.P., Champagne, B.J.: Virtual reality simulator training for
  laparoscopic colectomy: what metrics have construct validity?
\newblock Diseases of the Colon \& Rectum \textbf{57}(2), 210--214 (2014)

\bibitem{king2016immersive}
King, F., Jayender, J., Bhagavatula, S.K., Shyn, P.B., Pieper, S., Kapur, T.,
  Lasso, A., Fichtinger, G.: An immersive virtual reality environment for
  diagnostic imaging.
\newblock Journal of Medical Robotics Research \textbf{1}(01), 1640003 (2016).
\newblock \doi{10.1142/S2424905X16400031}

\bibitem{sousa2017vrrrroom}
Sousa, M., Mendes, D., Paulo, S., Matela, N., Jorge, J., Lopes, D.S.:
  {VRRRRoom: Virtual Reality for Radiologists in the Reading Room}.
\newblock In: Proceedings of the 2017 CHI Conference on Human Factors in
  Computing Systems, pp. 4057--4062. ACM (2017).
\newblock \doi{10.1145/3025453.3025566}

\bibitem{wirth2018evaluation}
Wirth, M., Gradl, S., Sembdner, J., Kuhrt, S., Eskofier, B.M.: Evaluation of
  interaction techniques for a virtual reality reading room in diagnostic
  radiology.
\newblock In: The 31st Annual ACM Symposium on User Interface Software and
  Technology, pp. 867--876. ACM (2018)

\bibitem{mirhosseini2019immersive}
Mirhosseini, S., Gutenko, I., Ojal, S., Marino, J., Kaufman, A.: Immersive
  virtual colonoscopy.
\newblock IEEE transactions on visualization and computer graphics
  \textbf{25}(5), 2011--2021 (2019)

\bibitem{randall2015oculus}
Randall, D., Metherall, P., Bardhan, K.D., Spencer, P., Gillott, R.,
  de~Noronha, R., Fenner, J.W.: The oculus rift virtual colonoscopy:
  introducing a new technology and initial impressions.
\newblock Journal of Biomedical Graphics and Computing \textbf{6}(1) (2015).
\newblock \doi{10.5430/jbgc.v6n1p34}

\bibitem{tciaCTC}
Smith, K., Clark, K., Bennett, W., Nolan, T., Kirby, J., Wolfsberger, M.,
  Moulton, J., Vendt, B., Freymann, J.: Data from {CT{\_}COLONOGRAPHY}.
\newblock Retrieved April 23, 2017
  \url{http://doi.org/10.7937/K9/TCIA.2015.NWTESAY1} (2015)

\bibitem{ribeiro20093}
Ribeiro, N., Fernandes, P., Lopes, D., Folgado, J., Fernandes, P.: 3-d solid
  and finite element modeling of biomechanical structures—a software
  pipeline.
\newblock In: 7th EUROMECH Solid Mechanics Conference (2009)

\bibitem{tagliasacchi2012mean}
Tagliasacchi, A., Alhashim, I., Olson, M., Zhang, H.: Mean curvature skeletons.
\newblock Computer Graphics Forum \textbf{31}(5), 1735--1744 (2012).
\newblock \doi{10.1111/j.1467-8659.2012.03178.x}

\bibitem{laviola2000discussion}
LaViola~Jr, J.J.: A discussion of cybersickness in virtual environments.
\newblock ACM SIGCHI Bulletin \textbf{32}(1), 47--56 (2000)

\end{thebibliography}

\end{document}